\documentclass[12pt]{iopart}
\usepackage{graphicx}
\begin{document}

\title[High-spin structures of $N=82$ isotones]{High-spin structures of $^{136}_{\ 54}$Xe, $^{137}_{\ 55}$Cs, $^{138}_{\ 56}$Ba, $^{139}_{\ 57}$La, and $^{140}_{\ 58}$Ce : A shell model description}
\author{P.C. Srivastava}
\address{Department of Physics, Indian Institute of Technology, Roorkee - 247667, India and}
\address{Instituto de Ciencias Nucleares, Universidad Nacional Aut\'onoma de M\'exico, 04510 M\'exico, D.F., Mexico}
\ead{pcsrifph@iitr.ernet.in}
\author{ M.J. Ermamatov}
\address{Instituto de Ciencias Nucleares, Universidad Nacional Aut\'onoma de M\'exico, 04510 M\'exico, D.F., Mexico and}
\address{Institute of Nuclear Physics, Ulughbek, Tashkent 100214, Uzbekistan}
\author{Irving O. Morales}
\address{Instituto de Ciencias Nucleares, Universidad Nacional Aut\'onoma de M\'exico, 04510 M\'exico, D.F., Mexico}

\begin{abstract}

In the present work recently available experimental data [A. Astier {\it et al}, Phys. Rev. C {\bf85}, 064316 (2012)] for high-spin states 
of five $N=82$ isotones, $^{136}_{\ 54}$Xe, $^{137}_{\ 55}$Cs, $^{138}_{\ 56}$Ba, $^{139}_{\ 57}$La, and $^{140}_{\ 58}$Ce have been interpreted with shell model calculations. The calculations have been performed in the 50-82 valence shell composed of  
$1g_{7/2}$, $2d_{5/2}$, $1h_{11/2}$, $3s_{1/2}$, and $2d_{3/2}$ orbitals. We have compared our results with the available experimental data for excitation energies including high-spin states, occupancy numbers and transition probabilities. As expected the structure of these isotones are due to proton excitations across $Z=50$ shell.  The structure of the positive-parity states are mainly from $(\pi g_{7/2}\pi d_{5/2})^n$ and 
$(\pi g_{7/2}\pi d_{5/2})^{n-2}(\pi h_{11/2})^2$ configurations, while the negative-parity states have $(\pi g_{7/2}\pi d_{5/2})^{n}(\pi h_{11/2})^1$ configuration. 
Additionally, for the $^{136}_{\ 54}$Xe, $^{137}_{\ 55}$Cs and $^{138}_{\ 56}$Ba isotones the excitation of the neutrons across $N=82$ gap is important.

\end{abstract} 

\submitto{\JPG}
\maketitle

\section{Introduction}

Shell structure near doubly magic nuclei is important 
for the study of weak interaction rates.  One of the $N=82$ isotones, $^{136}$Xe is a candidate of neutrinoless double beta decay, for determining neutrino
mass \cite{PhysRevLett.109.042501}.
For the $N=82$ isotones low and high-spin states produced by the spontaneous fission were reported in the literature \cite{Holt1997107,PhysRevC.80.044320,
PhysRevLett.77.3743,PhysRevC.59.3066}.
Recently, Astier {\it et al} \cite {PhysRevC.85.064316} populated the high-spin states of
$^{136}_{\ 54}$Xe, $^{137}_{\ 55}$Cs, $^{138}_{\ 56}$Ba, $^{139}_{\ 57}$La, and $^{140}_{\ 58}$Ce  
isotopes using fusion-fission reactions.

The aim of this work is to study $N=82$
isotones including several newly populated high-spin states  by Astier {\it et al} \cite {PhysRevC.85.064316} within framework of shell model. This would add more comprehensive information to the earlier shell model studies in Refs.~\cite{Holt1997107,PhysRevC.80.044320}.
The importance of $g_{7/2}$, $d_{5/2}$ and $h_{11/2}$ protons orbitals are crucial in this study. 
This can be seen from the evaluation of experimental values of the proton occupation number of the ground states for the different orbitals
in the even $N=82$ isotones in Fig. 1.   

Previously,  in this region we have analyzed experimentally observed slow $E3$ transition in $^{136}$Cs  at ISOLDE facility at CERN within framework of shell model~\cite{PhysRevC.84.014329}.  Shell model results on the spectroscopic properties of
$fp$ and $fpg$ shell nuclei were reported in our earlier works~\cite{sri09,Sri10,Sri12}.
The $B(E2)$ transitions trends in the tin isotopes using generalized seniority approach have recently been reported by I. O. Morales
{\it et al}~\cite{Morales2011606}.

\begin{figure}[h]
\hspace{1.5cm}\includegraphics[width=11.4cm]{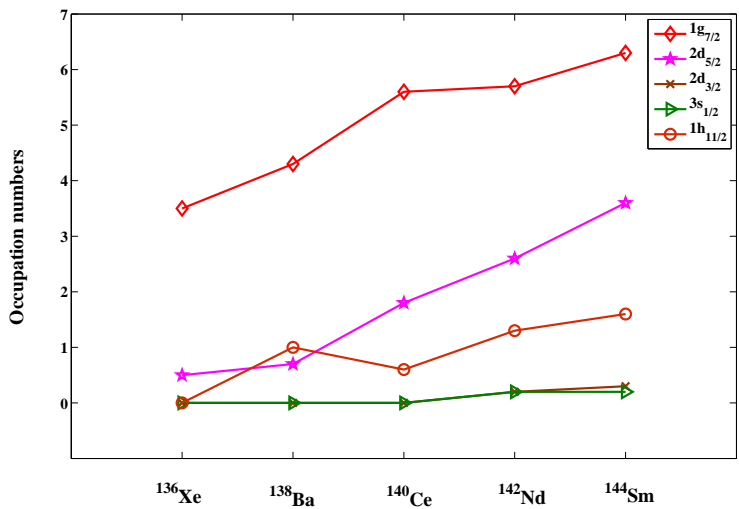}
\caption{
Variation of experimental occupation numbers \cite{PhysRevC.3.1199} for the ground states for even-mass $N=82$ isotones.}
\label{f_occu}
\end{figure}

This work is organized as follows:  comprehensive
comparison of shell-model results and experimental data is given in Section 2. In Section 3, 
transition probabilities are compared with the available experimental data.
Finally, concluding remarks are drawn in Section 4.
 
\section{Shell model results and discussions}

The shell-model calculations for the N=82 isotones have been
performed in the 50-82 valence shell composed of the orbits $1g_{7/2}$, $2d_{5/2}$,
$1h_{11/2}$, $3s_{1/2}$, and $2d_{3/2}$. The excited states are mainly
due to proton excitations. The full shell-model calculation we have performed with
SN100PN interaction due to Brown {\it et al}~\cite{PhysRevC.71.044317}. This interaction has four parts: 
neutron-neutron, neutron-proton, proton-proton and Coulomb repulsion between the protons.
The single-particle energies for the neutrons
are -10.610, -10.290, -8.717, -8.716, and -8.816 MeV
for the $1g_{7/2}$, $2d_{5/2}$, $2d_{3/2}$, $3s_{1/2}$, and $1h_{11/2}$ orbitals, respectively,
and those for the protons are 0.807, 1.562, 3.316, 3.224,
and 3.603 MeV~\cite{PhysRevC.84.014329}. The results shown in this work were obtained with the code NuShell \cite{MSU-NSCL}. In Ref. \cite{PhysRevC.85.064316} shell model results have been reported with empirical
interaction, while SN100PN interaction~\cite{PhysRevC.71.044317} is derived from the free nucleon-nucleon CD-Bonn potential. A new shell model Hamiltonian developed by  
Coraggio {\it et al} \cite{PhysRevC.80.044320} starting from a $V_{low-k}$ derived from the CD-Bonn potential, reported results for low-energy states for $N=82$ isotones. In the present study we also compare our results with 
previously available shell model results in Refs. \cite { PhysRevC.80.044320,PhysRevC.85.064316}.

\subsection{Analysis of spectra}

For the $N=82$ isotones proton excitations are important among the $1g_{7/2}$, $2d_{5/2}$, $2d_{3/2}$, $3s_{1/2}$, and $1h_{11/2}$ orbitals, above $Z=50$ shell closure. The neutrons will not contribute in the structure of these nuclei because of the $N=82$ shell closure.
 In this section we perform 
shell model calculations  for the chain of isotones $^{136}_{\ 54}$Xe, $^{137}_{\ 55}$Cs, $^{138}_{\ 56}$Ba, $^{139}_{\ 57}$La, and $^{140}_{\ 58}$Ce,
with valence protons in 50-82 shell in order to describe positive and negative-parity levels of these nuclei. In Table~\ref{ocup} the proton occupancy numbers calculated with shell model are given. They are  compared with the available experiment for the even isotones. For the odd isotones, we have listed predictions of the shell model. Results are in very good agreement with the available experimental data. There are 4 to 7 protons in average in $g_{7/2}$ and $d_{5/2}$
orbitals. Probability that one proton occupies $d_{3/2}$ orbital is small and even smaller for $s_{1/2}$ orbital, which will be seen also from the discussions below.  

\begin{table}[htbp]
\begin{center}
\caption{\label{ocup}Experimental~\cite{PhysRevC.3.1199} and calculated occupation 
numbers for the ground states of the $N=82$ isotones.}
\vspace{0.2cm}
\begin{tabular}{l|ccccc|c}
\hline
 &  $\pi$$g_{7/2}$ & $\pi$$d_{5/2}$ & $\pi$$h_{11/2}$ & $\pi$$d_{3/2}$ &
$\pi$$s_{1/2}$ & 
            $\pi$$g_{7/2} + \pi$$d_{5/2}$ \\
\hline
{\bf $^{136}$Xe} & & & & & & \\
Expt.        & 3.5(4) & 0.5(2) & 0.0(7) & 0.0(2) & 0.0(2) & 4.0(6) \\
SM         & 2.97   & 0.59   & 0.29   & 0.11   & 0.04   & 3.56 \\ 
\hline
{\bf $^{137}$Cs} & & & & & & \\
Expt.        & - & - & - & -& - & - \\
SM         & 3.78   & 0.75   & 0.30   & 0.12   & 0.04   & 4.53 \\ 
\hline
{\bf $^{138}$Ba} & & & & & & \\
Expt.        & 4.3(4) & 0.7(3) & 1.0(8) & 0.0(2) & 0.0(2) & 5.0(7) \\
SM         & 3.98   & 1.22   & 0.53   & 0.19   & 0.07   & 5.20 \\
\hline
{\bf $^{139}$La} & & & & & & \\
Expt.        & - & - & - & - & - & - \\
SM         & 4.56   & 1.58  & 0.56   & 0.21   & 0.09   & 6.14 \\
\hline
{\bf  $^{140}$Ce} & & & & & & \\
Expt.        & 5.6(3) & 1.8(2) & 0.6(4) & 0.0(2) & 0.0(2) & 7.4(5) \\
SM         & 4.75   & 2.00   & 0.82   & 0.29   & 0.13   & 6.75 \\
\hline
\end{tabular}
\label{tab:occupation}
\end{center}
\end{table}

\subsubsection{{\bf $^{136}_{~54}$Xe:}\label{Xe136}}

\begin{figure*}
\includegraphics[width=14.4cm]{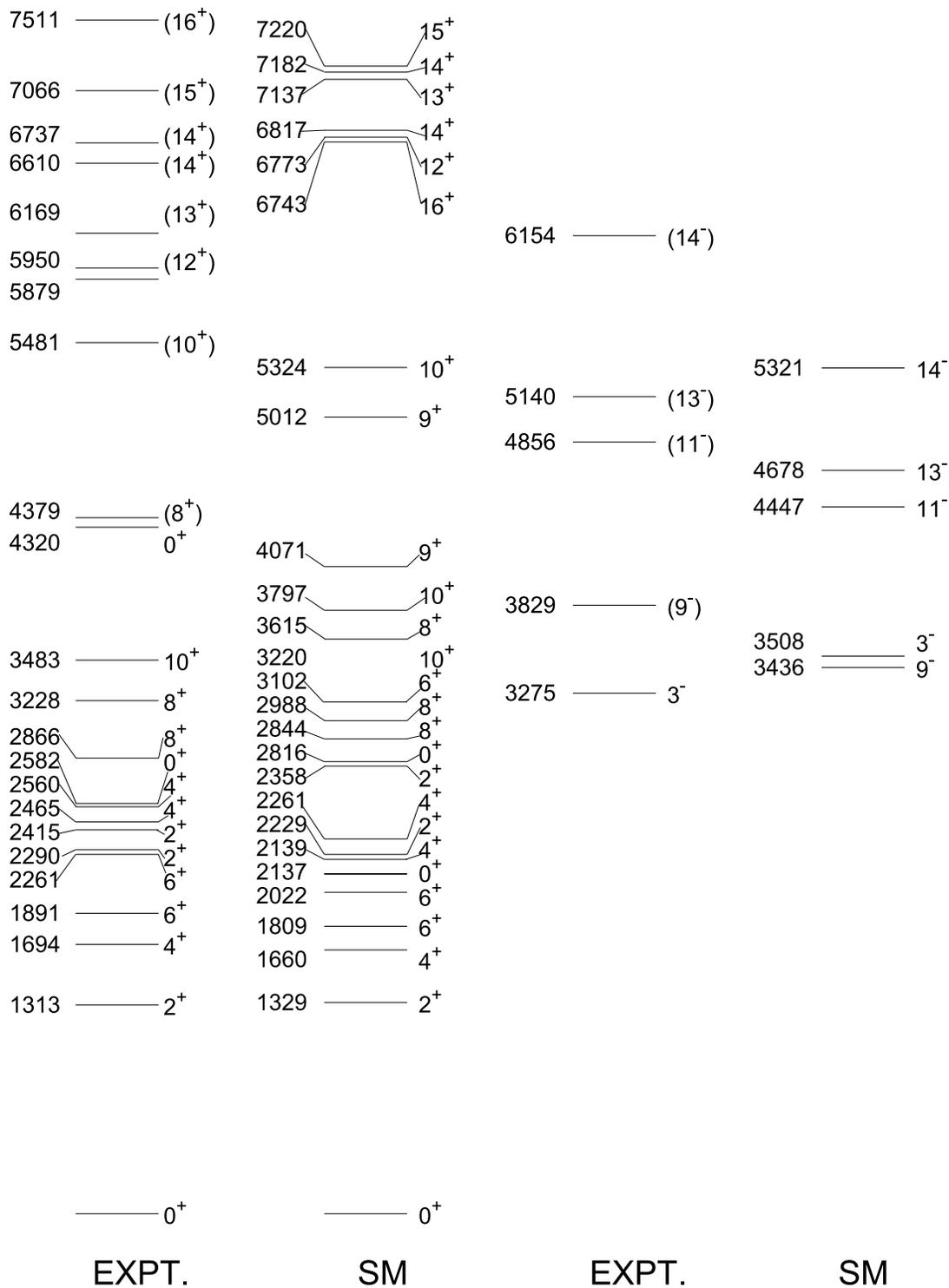}
\caption{
Comparison of experimental \cite{PhysRevC.85.064316,nndc} and calculated excitation spectra for $^{136}$Xe 
using SN100PN interaction.}
\label{f_xe136}
\end{figure*}

Comparison of the calculated values with the experimental data is shown in Fig.~\ref{f_xe136}. Recently, positive-parity levels of $^{136}$Xe  are extended up to 7.9 MeV in the experiment~\cite{PhysRevC.85.064316}. There are  also four experimental measured negative-parity levels.  

First four calculated positive-parity levels are in a very good agreement with the experiment. The $6_2^+$ is predicted by SM  239 keV lower than in the experiment. The experimentally observed $0^+_2$ level is much higher than the calculated $0^+_2$ level, however it is very close to the calculated $0^+_3$ level. The levels $2^+_2$, $2^+_3$, $4^+_2$ and $4^+_3$ at 2290, 2415, 2465 and 2560 keV are predicted by shell model at 2229, 2358, 2139, and 2261 keV, respectively. The predicted $8_1^+$  level differs from the experimental one by only 22 keV.
The difference increases up to 240 keV for the level $8^+_2$ at 3228 keV. The $10^+_1$ level at 3483 keV is predicted by shell model at 3220 keV. The $8^+_3$ is  more than 1 MeV lower in the SM calculation. According to shell model in the states 
$0^+_1$, $2^+_1$,$4^+_1$, $6^+_1$ and $8^+_1$  all four protons prefer to be in $g_{7/2}$ orbital with 53.3, 63.6, 68.8, 66.4 and 84.1\% probabilities, respectively.
The newly measured  level $10^+_2$ in~\cite{PhysRevC.85.064316} is very close to the $10^+_3$ in the calculation. There is the level sequence $10^+_2$, $9^+_1$ and $9^+_2$
levels between the levels $8^+_3$ and $10^+_3$ in the calculation which does not exist in the experiment. The rest higher spin states are more compressed 
in the calculation. The level $16^+_1$ is predicted much lower than in the experiment. SM predicts larger values for the $12^+_1$, $13^+_1$, $14^+_1$, 
$14^+_2$,  $15^+_1$, $16^+_1$ and $17^+_1$ levels.
 One proton promotion from $g_{7/2}$ to $d_{5/2}$ gives $10^+_1$ with 93.3\% probability. The structure of the  $12^+_1$, $14^+_1$ and $16^+_1$ states are connected with exciting two protons from $g_{7/2}$ to $h_{11/2}$ orbital with more than 90\% probabilities.   

Our  predicted results for $2_1^+ $, $4_1^+ $ and $6_1^+$ are close to the experiment in comparison to
\cite{PhysRevC.80.044320}.  As we move from   $^{136}$Xe to $^{140}$Ce the results for
$2_1^+$, $4_1^+$ and $6_1^+$ levels show better agreement in comparison to our results.
 The values for positive-parity sates up to $10_1^+$ are similar as in Ref.~\cite{PhysRevC.85.064316}.

For the negative-parities,  $3^-$ state is the lowest in the experiment while in calculation $9^-$ is the lowest one.
 Present shell model study 
predicts $14^-$ $\rightarrow$ $13^-$ $\rightarrow$ $11^-$ $\rightarrow$ $3^-$ $\rightarrow$ $9^-$ sequence for negative parity, while $3^-$ state is not shown in the shell model calculation of \cite{PhysRevC.85.064316}.
 Three protons are in $g_{7/2}$ and one proton is in $h_{11/2}$ for the negative-parity levels up to $13^-$ with 50.2, 81.4, 94.9 and 96.5\% probabilities, respectively. The probability of $g^2_{7/2}d^1_{5/2}h^1_{11/2}$ configuration is 99.7\% for the $14^-$ level.
With this model space we predict results up to $16^+$.

\subsubsection{{\bf $^{137}_{~55}$Cs:}\label{Cs137}}

\begin{figure*}
\includegraphics[width=14.4cm]{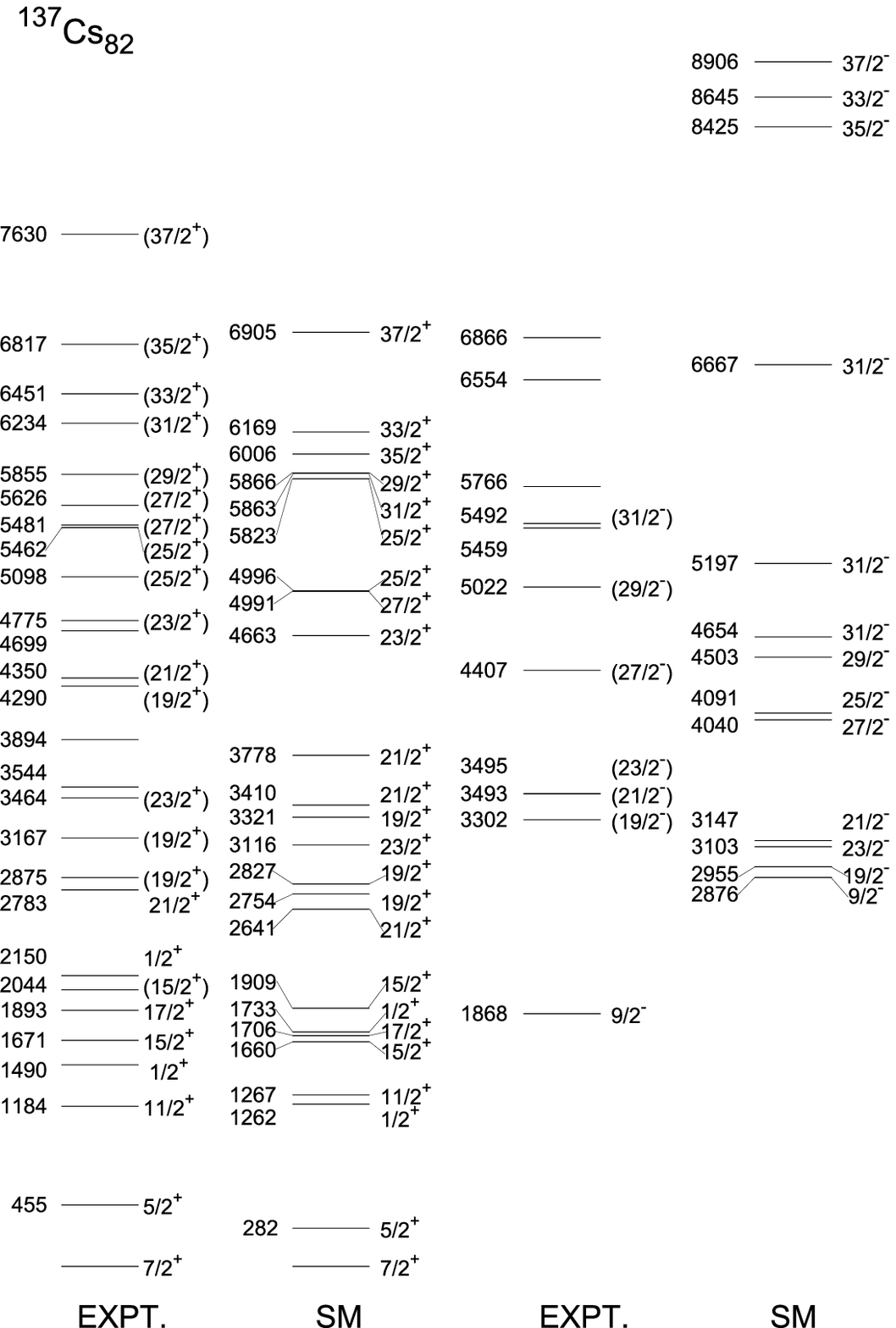}
\caption{Comparison of experimental \cite{PhysRevC.85.064316,nndc} and calculated excitation spectra for $^{137}$Cs 
using SN100PN interaction.}
\label{f_cs137}
\end{figure*}

The high-spin level scheme of the odd proton nucleus $^{137}$Cs
has been extended up to $\sim$7.6 MeV excitation energy and spin
(37/2+)~\cite{PhysRevC.85.064316}. As is seen from Fig.~\ref{f_cs137} only the sequence of  the levels $1/2^+_1$ and $1/2^+_2$ is different from the experiment up to $23/2^+_1$. Both of them are predicted lower than in the experiment. 
There are two levels, to which spins are not assigned in the experiment, between the levels $(23/2^+_1)$ and
$(19/2^+_3)$.  The next two levels ($19/2^+_3$) and ($21/2^+_2$) are predicted $\sim$1 MeV lower by the calculation. Spin is not assigned to the newly measured level at 4699 keV between the levels  $(21/2^+_2)$ and $(23/2^+_2)$. 
This may be  $21/2^+_3$  according to the shell model calculation.  The highest state $(37/2_1^+)$ is predicted by calculation 725 keV lower than in the experiment. The $7/2^+_1$, $9/2^+_1$, $11/2^+_1$ and $15/2^+_1$ and states are produced by 5 protons in $g_{7/2}$ orbital with
45.4, 48.3, 51.8 and 47.4\% probabilities, respectively. The  $1/2^+_1$, $5/2^+_1$, $13/2^+_1$, $17/2^+_1$, $19/2^+_1$ and $21/2^+_1$, states are due to one proton promotion  from $g_{7/2}$ to $d_{5/2}$ in the previous configuration with 40.7, 63.4, 80.4, 79.5, 86.9 and 92.7\%  probabilities, respectively. For the $1/2^+_1$ state also contribution of $g^{4}_{7/2}s^1_{1/2}$ configuration is essential: it is predicted by 23.6\% probability. The level $23/2^+_1$ is produced by two proton excitations from $g_{7/2}$ to $d_{5/2}$. Three and two protons are respectively in $g_{7/2}$ and $h_{11/2}$ orbitals in producing $17/2^+_1$, $25/2^+_1$, $27/2^+_1$, $29/2^+_1$, $31/2^+_1$, $33/2^+_1$ and $35/2^+_1$ levels.

The calculated $9/2^-$ level is $\sim$1 MeV higher than the experimental one.
The next triple of the negative-parity states are located very close to each other both in the experiment and calculation. The $21/1^-$ and 
$23/2^-$ states are interchanged with respect to the experiment. The  $27/2^-$ comes after around $\sim$ 1 MeV gap both in the experiment and calculation. The next calculated level is $25/2^-$ which is not measured in the experiment. The calculated $29/2^-$ and $31/2^-$levels  are located 
lower as compared to the experiment. The spacing between these two levels is less than the experimental one. We have also shown other calculated levels in order to compare
with the experimental tentative levels at 5459, 5766, 6554 and 6866 keV.  For the negative-parity levels $9/2^-$, $19/2^-$, $21/2^-$, $23/2^-$, $25/2^-$ and $27/2^-$ there are 3 protons in $g_{7/2}$ and 1 in the $h_{11/2}$ with 61.2, 83.3, 78.9, 77.8, 91.9 and 93.9\% probabilities. One proton excitation from $g_{7/2}$ to $d_{5/2}$ in the previous structure gives the $29/2^-$ and $31/2^-$ levels with 98.0 and 97.8\% probabilities, respectively. The levels $33/2^-$, $35/2^-$ and $37/2^-$ have configuration $g^{2}_{7/2}h^3_{11/2}$ with 99.3, 98.8 and 97.6\%  probabilities, respectively.  The $5/2^+_1$and $11/2_1^-$ yrast states are located lower than in the experiment in this calculation, while they are higher  in \cite{PhysRevC.80.044320}. But the trends in both shell model results from  $^{137}$Cs to
 $^{139}$La are the same.

\subsubsection{{\bf $^{138}_{~56}$Ba:}\label{Ba138}}

\begin{figure*}
\hspace{1cm}\includegraphics[width=14.4cm]{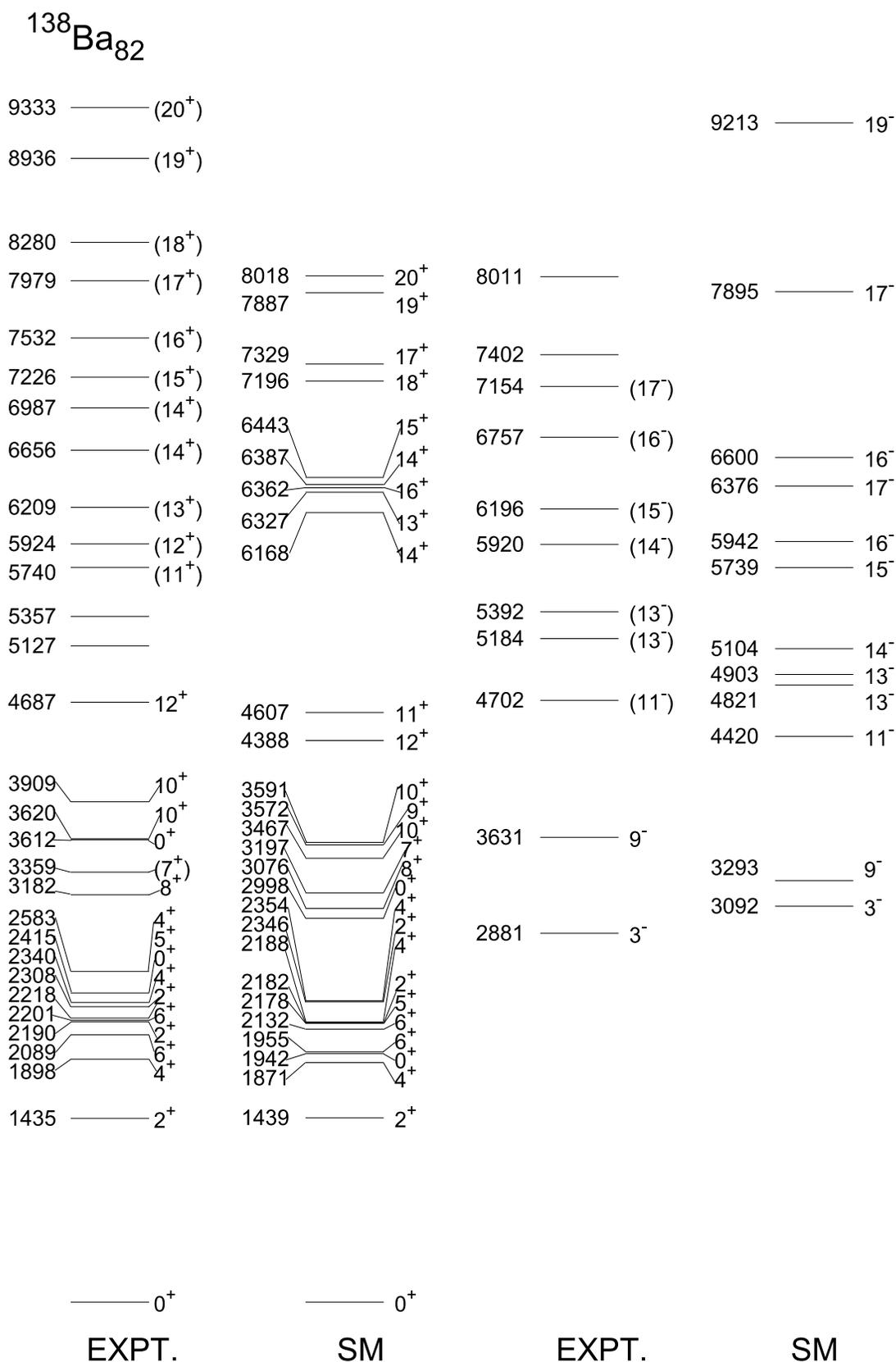}
\caption{ Comparison of experimental \cite{PhysRevC.85.064316,nndc} and calculated excitation spectra for $^{138}$Ba 
using SN100PN interaction.}
\label{f_ba138}
\end{figure*}

Experimental positive-parity levels are available up to $\sim$9 MeV excitation energy and spin $20^+$ for this nucleus~\cite{PhysRevC.85.064316}. 
Comparison of the calculated levels with the experimental data is presented in Fig.~\ref{f_ba138}. We have compared three eigenvalues with the experiment  up to $10^+$. The $2^+_1$ and $4^+_1$ are predicted by the calculation only with 4 and 27 keV difference, respectively. In the calculation there is $0^+_2$ at 1942 keV between $4^+_1$ and $6^+_1$ which is at 2340 keV in the experiment. 
 The difference between the experiment and calculation increases up to $\sim1.3$ MeV when reaching $20^+_1$ state.
The  $0^+_1$, $2^+_1$, $4^+_1$, $10^+_1$ and $12^+_1$ levels have $g_{7/2}^4d_{5/2}^2$ configuration with 34.6, 40.0, 34.2, 48.5, and 91.7\% probabilities.  There are five protons  in $g_{7/2}$ and one in $d_{5/2}$ orbital for the $6^+_1$ and $8^+_1$ levels with 42.6 and 54.8\% probabilities, respectively. One proton promotion from the previous configuration to $d_{3/2}$ orbital gives $13^+_1$ state with 97.3\% probability. The $14^+_1$,
$15^+_1$, $16^+_1$ and $17^+_1$ levels are due to the $g_{7/2}^4h_{11/2}^2$ configuration with 87.2, 86.1, 81.5 and 94.1\% probabilities, respectively.   In case of  $11^+$ state to the $14_1^+$, neutron excitation across the $N=82$ gap is important. This is also concluded in Ref. ~\cite{PhysRevC.85.064316}.
  
The calculated negative-parity levels we have shown up to $19^-_1$ in order to compare results also with experimentally tentative levels at 7402 and 8011 keV beyond $17^-$.  
It can be noted that as we move from $^{136}{\rm Xe}$ to $^{138}{\rm Ba}$ the calculation now predicts correct order of the $3^-$ and $9^-$ 
as in the experiment. This is due to the change of configuration from $g^3_{7/2}h^1_{11/2}$ to $g^4_{7/2}d^1_{5/2}h^1_{11/2}$.
The spacing between these levels are still less than in the experiment.
 Present shell model study also predicts as in Ref. ~\cite{PhysRevC.85.064316} that two $13^-$ states are very close as is observed in the experiment.

Negative-parity states are produced because of the odd number of protons in $h_{11/2}$ orbital. The $9^-$, $11^-$ and $13^-$ levels are due to $g^5_{7/2}h^1_{11/2}$ configuration with 50.6, 62.8, 65.8 and 81.3\% probabilities.  The promotion of one proton from $g_{7/2}$ to $d_{5/2}$ in the previous configuration gives the $3^-$, $14^-$ and $16^-$ levels with 60.2, 81.3 and 94.6\% probabilities.
For $17^-_1$ now there is one more proton excitation from $g_{7/2}$ to $d_{5/2}$. In $19/2^-$ state three protons are both in $g_{7/2}$ and $h_{11/2}$ orbital with 96.7\% probability. 

\subsubsection{{\bf $^{139}_{~57}$La}:\label{La139}}

\begin{figure*}
\includegraphics[width=14.4cm]{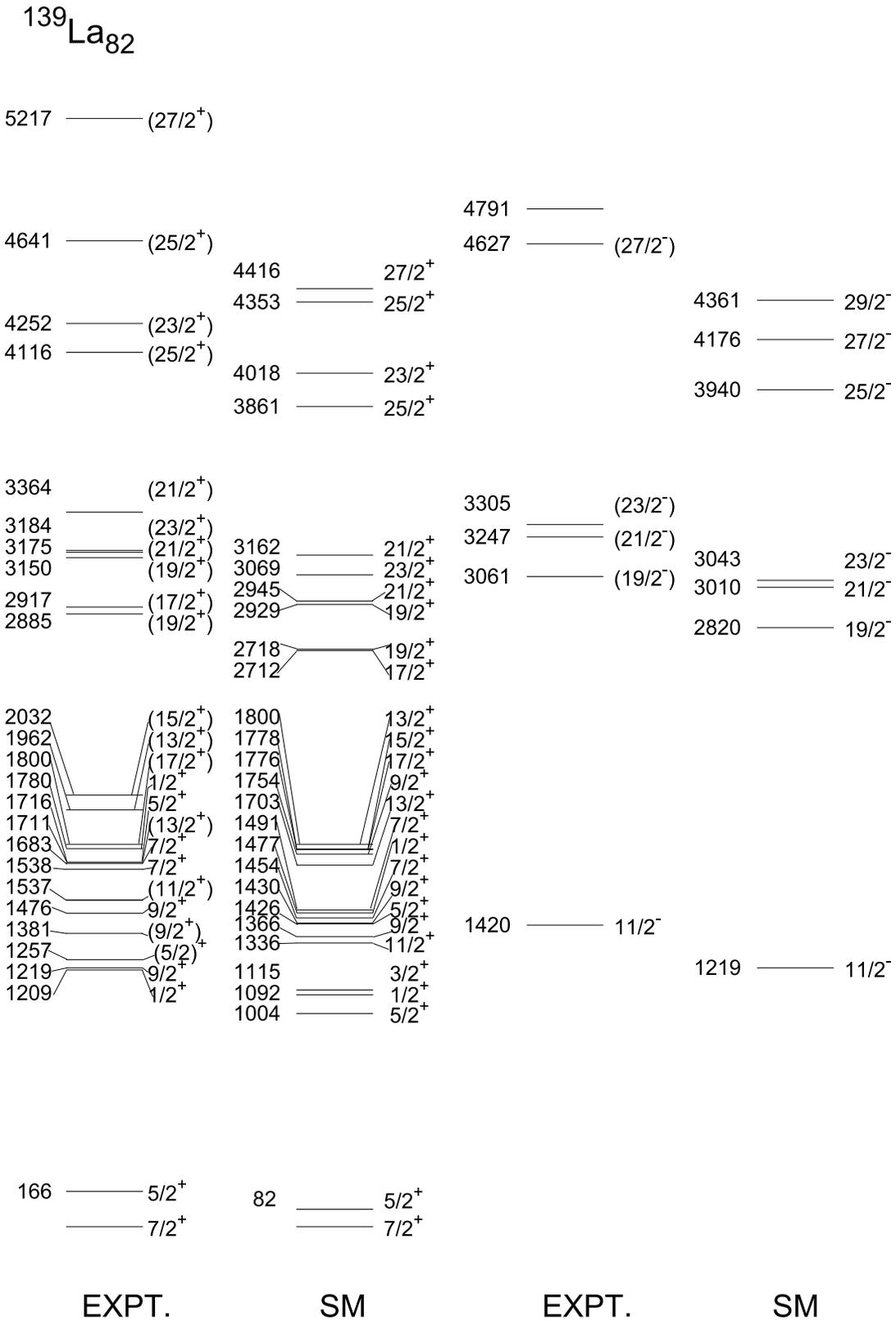}
\caption{ Comparison of experimental \cite{PhysRevC.85.064316,nndc} and calculated excitation spectra for $^{139}$La 
using SN100PN interaction.}
\label{f_la139}
\end{figure*}

The calculated positive- and negative-parity levels of $^{139}$La are given in Fig.~\ref{f_la139}. All calculated levels are lower as compared to the experimental ones. The first calculated $5/2^+$ is closely located
to the ground state as compared to the experimental one. The experimental levels from $1/2^+$ to $17/2^+$ are located between 1209 and 2032 keV energy range, while calculated energies are between 1004 and 1800 keV. The levels from $17/2^+$ to $21/2^+$ come after $\sim$900 keV gap in both experiment and calculation. The next sequence of the levels $23/2^+$ to $27/2^+$ are located after $\sim$800 keV gap both in the experiment and calculation.  The sequence of these levels are the same both in the experiment and calculation. For this nucleus $7/2^+_1$, $11/2^+_1$, $15/2^+_1$, $19/2^+_1$, $23/2^+_1$ and $27/2^+_1$ states are due to five protons in $g_{7/2}$ and remained two protons in $d_{5/2}$ for the first five states and in $h_{11/2}$ for the last state with 38.6, 51.6, 46.9, 79.5, 82.5 and 60.6\% probabilities, respectively.  
The $5/2^+_1$, $9/2^+_1$, $13/2^+_1$ and $17/2^-_1$ states have $g_{7/2}^6h_{11/2}^1$ configuration with 32.9, 34.2, 41.7 and 37.8\% probabilities.
There are four and three protons in $g_{7/2}$ and $d_{3/2}$ orbitals for the $21/2^+_1$ and  $25/2^+_1$ states with 70.7 and 92.9\% probabilities, respectively. $1/2^+_1$ state is due to $g_{7/2}^4d_{5/2}^2s_{1/2}^1$ configuration with 23.8\% probability.   In Ref.~\cite{PhysRevC.85.064316} the $15/2^+_1$
state is predicted to lie 20 keV below the $17/2^+_1$ state, hence the latter would not exhibit any delayed decay.
In the present study our predicted $15/2^+_1$ is higher than $17/2^+_1$, hence it explains correctly experimental
findings. It is also important to mention here that in the work of Astier {\it et al}  \cite{PhysRevC.85.064316}, the shell-model results for this isotope show overall better agreement with the experimental data in comparison to our results. Thus, for heavier $N=82$ isotones the SN100PN interaction
needs further tuning.

The calculated negative-parity levels are in reasonable agreement with the experiment. For the all negative-parity levels one proton is in $h_{11/2}$. For the $11/2^-$ and $19/2^-$ levels, the remained six protons are in $g_{7/2}$ with 30.9 and 41.1\% probabilities, respectively. 
Further, the $g_{7/2}^4d_{5/2}^1h_{11/2}^1$ configuration gives the structure for the $21/2^-$, $25/2^-$ and $27/2^-$ levels with 39.0, 77.8 and 49.3\% probabilities, respectively.  The $23/2^-$ is connected with the configuration where one more proton is moved from $g_{7/2}$ to $d_{5/2}$ in the previous configuration with 38.9\% probability. 

\subsubsection{{\bf $^{140}_{~58}$Ce:}\label{Ce140}}

\begin{figure*}
\includegraphics[width=14.4cm]{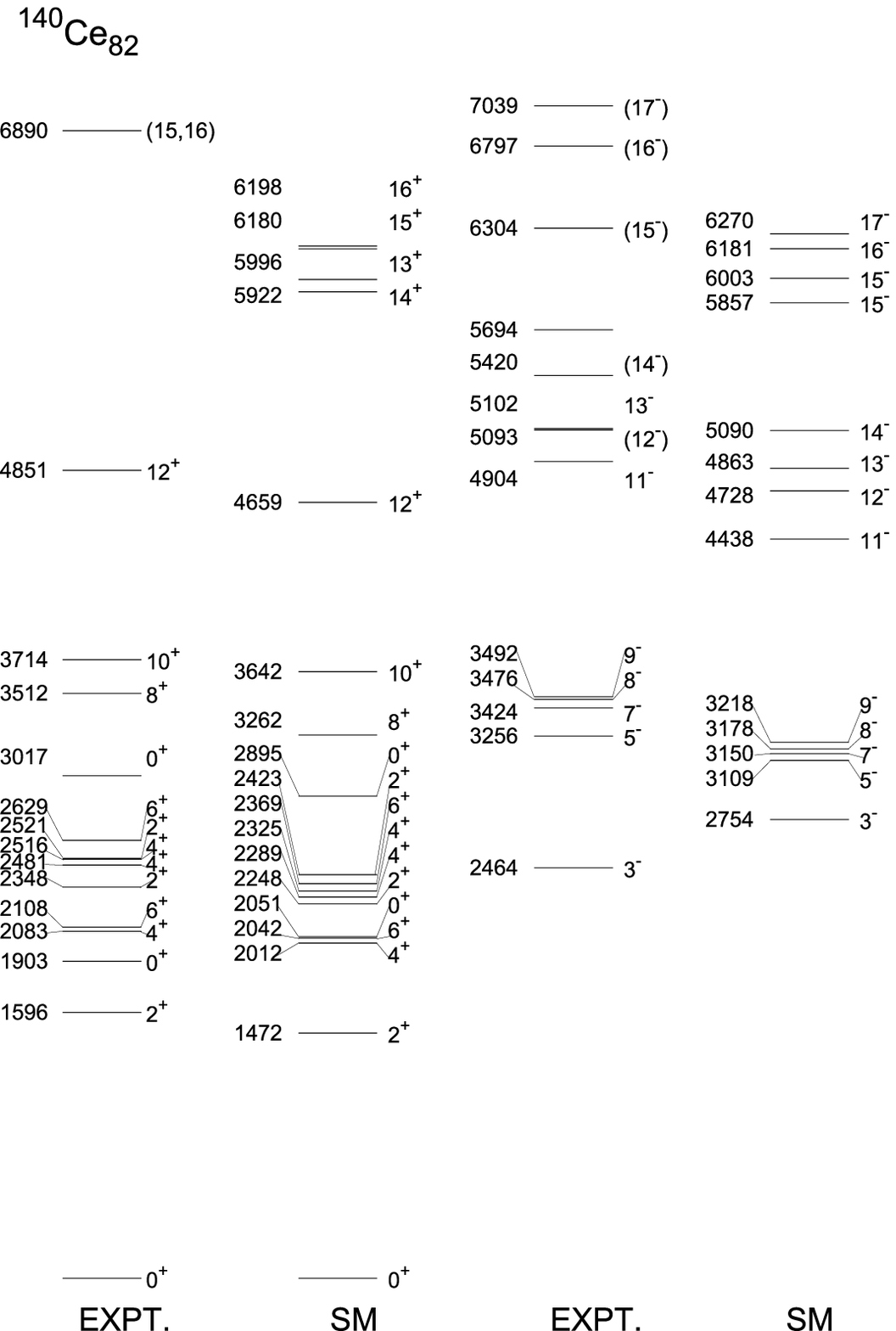}
\caption{Comparison of experimental \cite{PhysRevC.85.064316,nndc} and calculated excitation spectra for $^{140}$Ce 
using SN100PN interaction.}
\label{f_ce140}
\end{figure*}

The calculated positive- and negative-parity levels of $^{140}$Ce are given in Fig.~\ref{f_ce140}.
For the positive-parity levels comparison with the experiment is shown up to $16^+$. The calculated $0^+_2$ state is higher in the calculation. The other levels are located lower than in the experiment. This nucleus has eight protons outside the core. In the $2^+_1$, $4^+_1$, $6^+_1$, $10^+_1$, $13^+_1$, $14^+_1$, $15^+_1$ and $16^+_1$ states six protons are in $g_{7/2}$ and two are in $d_{5/2}$ for the first four states and in $h_{11/2}$ for the remained four states with 27.3, 28.6, 23.0, 46.1, 47.5, 50.4, 47.2 and 43.3\% probabilities, respectively. 
For the $6^+_1$, $8^+_1$ and $12^+_1$  states five protons are in $g_{7/2}$ and three protons are in $d_{5/2}$ with 38.1, 57.3,82.1\% probabilities, respectively. The  $17^+_1$ level has $g^5_{7/2}d^1_{5/2}h^2_{11/2}$ configuration with 70.3\% probability.

The calculated negative-parity levels show the same sequence as in the experiment. We have given two calculated eigenvalues for $15^-$ level in order to compare with the experimental level at 5697 keV to which spin is not assigned.   All the calculated levels are predicted lower than in the experiment. The negative-parity levels are produced following three configurations: $g_{7/2}^6d_{5/2}^1h_{11/2}^1$ for the $3^-$, $5^-$, $13^-$ and $14^-$ states with
38.0, 24.4, 45.6 and 40.5\% probabilities, respectively,  $g_{7/2}^5d_{5/2}^2h_{11/2}^1$ for the $7^-$, $8^-$, $9^-$,  $11^-$, $15^-$ and $17^-$ states with, 28.4, 40.4, 40.6, 52.4, 73.2 and 79.6\% probabilities, respectively and $g_{7/2}^4d_{5/2}^3h_{11/2}^1$  for the $16^-$ state with 60.3\% probability.

It is important to study the configuration across $N=82$ gap with neutron excitations.
Because from the above discussions it can be seen for $^{136}$Xe the 
 experimentally observed state at 7946 keV is too low in energy to come from the $\pi (g_{7/2}^2 d_{5/2}^2)$ configuration and  for  the $^{138}$Ba
 the intermediate high-spin states beyond the $11^+$ level,  the calculated results are not in a good agreement with the experiment.  In $^{137}_{~55}$Cs, the configuration of the highest-spin states is mainly from $(\pi g_{7/2}\pi d_{5/2})^{3}(\pi h_{11/2})^2$. For this nucleus, in case of intermediate-spin, the $(\pi g_{7/2}\pi d_{5/2})^{5}(\nu h_{11/2})^{-1}(\nu f_{7/2})^{+1}$ configuration is also important.
It can be noted that the difference between ($2^+_1$) and the ground state ($0^+$) energies of even isotones starting from $^{136}$Xe to $^{140}$Ce shows that it is  increased by the increasing of the proton number outside $Z=50$ closed shell in both calculation and experiment. This is an indication of persistence of $N=82$ magic number for neutrons. In the Ref.~\cite{PhysRevC.85.064316}, the negative parity states shown 
from $9_1^-$ onwards, while in the present study apart from correctly reproducing sequence of levels from $3_1^-$ up to $17_1^-$ the spacing between levels also show the same trend as in the experiment.  For this isotope yrast $2_1^+$, $4_1^+$ and $6_1^+$  levels show better agreement with the experimental data  in comparison to Ref.~\cite { PhysRevC.80.044320}.

\begin{table}[hbtp]
\caption{ Experimental and calculated $B(E2)$ and $B(E3)$  in W.u. for different
transitions.}
\label{t_be2}
\begin{center}
\begin{tabular}{rccc}
\hline
Nucleus  & ~~Transition & ~~Expt. &~~ Calc. \\        
\hline
$^{136}$Xe &B(E2;$2^{+}_{1} \rightarrow 0^{+}_{1}$)\hspace{1.cm} & 9.7 (4) ref.\cite{nndc1}& 7.4 \\ 
           &B(E2;$4^{+}_{1} \rightarrow 2^{+}_{1}$)\hspace{1.cm} & 1.281 (17)  & 0.9 \\
           &B(E2;$6^{+}_{1} \rightarrow 4^{+}_{1}$)\hspace{1.cm} & 0.0132 (4) &
0.089  \\
\hline
$^{137}$Cs &B(E2;$5/2^{+}_{1} \rightarrow 7/2^{+}_{1}$)\hspace{1.cm} & $>$6.8 & 0.02
 \\ 
\hline
$^{138}$Ba &B(E2;$2^{+}_{1} \rightarrow 0^{+}_{1}$)\hspace{1.cm} & 10.8 (5)& 11.4   
\\ 
           &B(E2;$4^{+}_{1} \rightarrow 2^{+}_{1}$)\hspace{1.cm} & 0.2873(15) & 0.20
   \\
           &B(E2;$6^{+}_{1} \rightarrow 4^{+}_{1}$)\hspace{1.cm} & 0.053(7) & 0.21  
 \\
           &B(E2;$2^{+}_{2} \rightarrow 0^{+}_{1}$)\hspace{1.cm} & 2.01(20) & 0.72  
 \\
           &B(E2;$4^{+}_{2} \rightarrow 6^{+}_{1}$)\hspace{1.cm} & 2.0(9) & 1.1    \\
\hline
$^{139}$La &B(E2;$9/2^{+}_{1} \rightarrow 5/2^{+}_{1}$)\hspace{1.cm} & 1.79(24) &
9.6   \\ 
           &B(E2;$7/2^{+}_{2} \rightarrow 7/2^{+}_{1}$)\hspace{1.cm} & 11.9(16) &
2.13   \\ 
           &B(E2;$5/2^{+}_{2} \rightarrow 7/2^{+}_{1}$)\hspace{1.cm} & 19.3(22) &
17.8   \\          
\hline
$^{140}$Ce &B(E2;$2^{+}_{1} \rightarrow 0^{+}_{1}$)\hspace{1.cm} & 13.8(3) & 14.9   
\\ 
           &B(E2;$0^{+}_{2} \rightarrow 2^{+}_{1}$)\hspace{1.cm} & 11.5(9) & 6.0    
\\ 
           &B(E2;$4^{+}_{1} \rightarrow 2^{+}_{1}$)\hspace{1.cm} & 0.137(4) & 0.11  
 \\ 
           &B(E2;$6^{+}_{1} \rightarrow 4^{+}_{1}$)\hspace{1.cm} & 0.29(6) & 0.15   \\
           &B(E2;$10^{+}_{1} \rightarrow 8^{+}_{1}$)\hspace{1.cm} & 0.46(13) & 0.21 \\
           &B(E3;$3^{-}_{1} \rightarrow 0^{+}_{1}$)\hspace{1.cm} & 27(6) & 9.3 \\
\hline
\end{tabular}
\label{be2}
\end{center}
\end{table}
\section{Transition probability analysis}
The comparison of the transition probabilities with the experiment is given in Table~\ref{t_be2}. The effective charge $e_p$=1.5 is used in the calculation. The  calculated values of $E2$ transition probabilities are in excellent agreement with the experimental ones for the even isotones. The only $E2$ transition existing in the experiment is measured to be more than 6.8 W.u. for the $^{137}$Cs. For this transition too weak $E2$ transition is predicted by shell  model. Calculations are in reasonable agreement also for the $^{139}$La isotone. The value of $E2$ transition probability from $9/2_1^+$ to $5/2_1^+$ is predicted larger and that of from $7/2_2^+$ to $7/2_1^+$ is smaller as compared to the experiment. Very good agreement of the last transition is seen from the Table. We have shown in Table~\ref{t_be2} also comparison of one calculated value  of $E3$ transition probability with its experimentally available counterpart for the $^{140}$Ce.  Our calculated B(E2; $2^{+}_{1} \rightarrow 0^{+}_{1}$) values for $^{136}$Xe and $^{138}$Ba are close to values which are reported in Ref. \cite{PhysRevC.80.044320}.

\section{Summary}

Our present work is motivated by recently available
high-spin state experimental data of five $N=82$ isotones, $^{136}_{\ 54}$Xe,
$^{137}_{\ 55}$Cs, $^{138}_{\ 56}$Ba,
 $^{139}_{\ 57}$La, and $^{140}_{\ 58}$Ce
for which we performed shell model calculations. This work adds
more information to Refs. \cite{Holt1997107,PhysRevC.80.044320,PhysRevC.85.064316}
where shell model results for even $N=82$ isotones have been reported.
We can summarize our results by following:
\begin{itemize} 
\item The yrast states of the five $N=82$ isotones are very well described by shell model.

\item The results predicted by shell model for the occupancy numbers are in a very good agreement with the experimental data. The predicted 
numbers also show the predominance of the $\pi g_{7/2}$, $\pi d_{5/2}$ and $\pi h_{11/2}$ orbitals. 
The  present results for the occupancy numbers 
are close to the experiment in comparison to the previous shell model results.

\item The structure for the positive-parity states of lighter isotones are from $(\pi g_{7/2}\pi d_{5/2})^n$
configuration. It is found that the protons prefer to fill the $\pi h_{11/2}$ orbital rather than $d_{3/2}$ and $s_{1/2}$ orbitals ($(\pi g_{7/2}\pi d_{5/2})^{n-2}(\pi h_{11/2})^2$ configuration) in the case of heavier isotones.  
The negative parity states are due to odd number of protons in $\pi h_{11/2}$ orbital.


\item The transition probabilities are in a good agreement with the experimental data and they are useful for understanding the structure of these isotones.

\item The neutron excitations across $N=82$ gap is important for the intermediate-high-spin states in case of 
    $^{136}_{\ 54}$Xe, $^{137}_{\ 55}$Cs, and $^{138}_{\ 56}$Ba isotones. Further theoretical development is needed 
   by constructing a new effective interaction to enlarge the model space which at least include $\nu f_{7/2}$ orbital from upper
    $hfpi$ shell.
\end{itemize}  

{\it Acknowledgement:} This work was supported in part by Conacyt, M\'exico, and by DGAPA, UNAM project IN103212.
MJE acknowledges support from grant No. 17901 of CONACyT projects CB2010/155633 and F2-FA-F177 of Uzbekistan
Academy of Sciences. One of the authors (P.C.S.) would like to thank to Prof. B. A. Brown for useful discussions.
Thanks are due to Dr. L. Fortunato for useful comments.


\section*{References}
\bibliographystyle{jphysg}

\end{document}